\documentclass[preprint,preprintnumbers,showpacs,prl]{revtex4}
\usepackage[english]{babel}
\usepackage{amsmath}

\usepackage{latexsym}

\usepackage{epsf}

\newcommand{\be}{\begin{equation}}
\newcommand{\ee}{\end{equation}}
\newcommand{\bea}{\begin{eqnarray}}
\newcommand{\eea}{\end{eqnarray}}

\newcommand{\hh}{\tilde{h}}

\newcommand{\cH}{{\cal H}}
\newcommand{\clE}{{\cal E}}
\newcommand{\prt}{\partial}
\newcommand{\vep}{\varepsilon}

\newcommand{\rgl}{\rangle}
\newcommand{\lgl}{\langle}
\newcommand{\sinc}{\mbox{\rm sinc}\,}
\begin{document}

\title{Wave localization as a manifestation of ray chaos
in underwater acoustics}

\author{A. Iomin and Yu. Bliokh }
\affiliation{Department of Physics, Technion, Haifa 32000, Israel}
\date{\today}

\begin{abstract}

Wave chaos is demonstrated by studying a wave propagation in a
periodically corrugated wave-guide. In the limit of a short wave
approximation (SWA) the underlying description is related to the chaotic
ray dynamics. In this case the control parameter of the problem is
characterized by the corrugation amplitude and the SWA parameter.
The considered model is fairly suitable and tractable for the analytical
analysis of a wave localization length.
The number of eigenmodes characterized the width of the localized wave
packet is estimated analytically.

\end{abstract}

 \maketitle

\section{Introduction}

Wave chaos phenomena reveal themselves in underwater acoustics.
It is known that a wave propagation with the wavelength $\lambda$ in a long
range--dependent wave-guide can be described under certain conditions
by the para\-bo\-lic equation in the limit of a small-angle propagation
\cite{tappert,jkps94}.
This equation  corresponds formally  to the Schr\"odinger equation
with an effective semiclassical parameter being of the order of
$\lambda/L$,  where $ L $ is the characteristic size of inhomogeneities.
It has been shown that this issue of quantum chaos is relevant to
the phenomenon of long--range acoustic wave propagation in the ocean,
where a ray-based approach is widely used in schemes of ocean acoustic
monitoring \cite{vz00,mw79,atoc,8a}. There is numerical evidence
\cite{pbtb88,sbt92,sfw97} that the phenomenon of ray chaos plays an
important role in long--range sound transition. Wave propagation in
the inhomogeneous wave-guide media can be considered from
the point of view of nonlinear Hamiltonian dynamics, where the
consideration can be reduced to a corresponding problem of ray dynamics
\cite{az91}. One of the most interesting features of ray dynamics is the
possibility of the dynamical chaos of rays when a wave-guide carries
longitudinalperiodic inhomogeneities. Unlike ray dynamics, where wave
features are omitted, the important specific features of wave propagation
can be taken into account by considering the parabolic equation for a
small--angle--propagation \cite{vz00,sz99}. The relation between the
wave equation and the corresponding ray dynamics was the issue of
extensive studies \cite{8a,az91,sz99,svz01}.

In this paper, we consider a model of the wave propagation in a
range--dependent wave-guide.
When $\lambda/L\rightarrow 0$, {\em i.e.} in the short wave approximation
(SWA),  the underlying  description can be reduced to the ray dynamics
which can be  chaotic due to range-dependent perturbations.
The nature of this chaos is similar to the chaotic billiards
\cite{gutzwiller,12a,az91}. The formulation of the problem for the wave
transmission in the framework of geometrical optics \cite{12b} is relevant
to the ocean monitoring problem where the parameter $ \lambda/L $ is small:
$ \lambda/L\sim 10^{-3}$ -- $ 10^{-2} $, and an accuracy of the SWA
seems fairly high \cite{8a}. However, a principal difference
between wave-like and ray-like descriptions, which may be negligible for
the stable ray dynamics, has a strong impact in the case of
chaotic rays. A manifestation of ray chaos in waves dynamics is
similar to the effects of quantum chaos. We can indicate at least two
such important manifestations: a breaking length $ x_{\lambda} $ along the
range--dependent (longitudinal) direction $x$, and a localization length
which characterizes the cut-off of the number of eigenmodes
$ \Delta n_{\lambda} $ of the propagated wave packet along $x$.
The first one, $x_{\lambda}$,
is similar to the breaking time $\tau_{\hbar}$ in quantum chaos
\cite{bz1978,12a}, sometimes called the Ehrenfest time. It shows an
accumulation of small wave effects due to the chaos of rays, which leads
to the breaking of the SWA for $x>x_{\lambda}$.
The second manifestation  of chaos related to localization of a wave packet
in the momentum space and the corresponding cut-off of the number of
eginmodes $\Delta n_{\lambda}$. This second manifestation  of chaos is
dynamical localization similar to the Anderson localization in solid
state.

The goal of this paper is to demonstrate the second manifestation
of wave chaos using a fairly convenient model of wave/ray propagation in a
periodically corrugated wave--guide (Fig. 1). The control parameter
of the problem can be written as a product \cite{bz,12a}
\begin{equation}\label{1a}
\vep/L=\left(2\vep/\lambda\right)\cdot\left(\lambda/2L\right),
\end{equation}
where the first multiplier characterizes the dimensionless amplitude of
the perturbation (corrugation), while the second one is the SWA parameter.
The problem, presented in the paper, is tractable for the analytical
analysis, and it  makes possible the calculate of the number of the
localized wave modes $\Delta n_{\lambda}$ for the two limiting cases
both $ \lambda\gg\vep $ and $ \vep\gg\lambda $.

\section{The semiclassical parameter}

We  first define the semiclassical parameter. Let us specify the general
properties of solutions of a 2D wave
equation in a homogeneous media bounded by two reflecting surfaces
$ y=0 ~~~\mbox{and} ~~~y=y_0+\vep\phi(x) $. Here  $\phi(x)$
is a corrugation function with the period $2\pi/k_0$, and the amplitude
of corrugation is $\vep$, while $ y_0\equiv L $ is an average
transverse size (or depth) of the channel (see Fig. 1).
In what follows, we also consider that a small parameter is $ \vep/y_0\ll
1 $. The notation $y_0$ is chosen to stress that the characteristic size has
the transversal direction. The solution of the wave equation can be
obtained in the following way. Assume that the time dependence of a wave
field is harmonic,
\[  \tilde{\Psi}(x,y,t)=\exp(-i\omega t)\Psi(x,y). \]
Applying conformal transformation to map the domain
$ x\in(-\infty,\infty),~y\in[0,y_0+\vep\phi(x)] $ to  the domain
$ x'\in(-\infty,\infty),~y'\in[0,y_0] $ and omitting primes for shortness,
one can write down in the first order of $ \vep/y_0 $ the stationary wave
equation in the form:
\be\label{jan3}
\Delta\Psi+\frac{\omega^2}{c_s^2}\left(1+\frac{\vep}{y_0}A(x,y)
\right)\Psi=0,
\ee
where $\Delta$ is the 2D Laplace operator, and $\omega$ is the frequency of
a sound wave with the velocity $c_s$.
It is obvious that the coefficient $ A(x,y) $ is a periodic function of $x$
with the period of the corrugation function. Thus, due to the Floquet
theorem, the solution is cast in the form
\be\label{jan4}
\Psi(x,y)=e^{ikx}B(x,y)\, ,
\ee
where $B(x,y)$ satisfies to the periodic $ B(x+2\pi/k_0,y)=B(x,y) $
and, e.g., Dirichlet boundary conditions, $ B(x,0)=B(x,y_0)=0 $.
In the zero order approximation, we obtain from (\ref{jan3}) the following
dispersion relation
\be\label{jan6}
\omega^2\equiv\omega_{m}^2(k)=c_s^2[k_m^2+p_m^2].
\ee
Here $ p_m^2\sim(\pi/y_0)^2m^2$ is the discrete spectrum of transverse wave
numbers (as a result of the boundary conditions) with the
corresponding eigenfunctions $ B(x,y)=b_{m}(x,y) $, where $ m=0,1,\dots $
\cite{note1}.
The dispersion relation for the fixed $ \omega $ is a constraint for
both the longitudinal $k_m$ and the transversal $p_m$ wave numbers.
Therefore the solution (\ref{jan4}) is a finite sum on $ m $ in a range
$ 0<m<m_{max} $:
\be\label{jan5}
\Psi(x,y)=\sum_{m=0}^{m_{max}}e^{ik_mx}b_m(x,y)\, .
\ee
 Since $ p_m\sim\pi m/y_0 $ for any type of the
boundary conditions and the maximal momentum is $ p_{max}=\omega/c_s $, we
obtain that
\be\label{jan8}
m_{max}\sim \omega y_0/\pi c_s=2y_0/\lambda .
\ee
Hence the validity of the geometrical optics limit (or
ray dynamics) is determined by the condition $ m_{max}=2y_0/\lambda\gg 1 $.
In this approximation, an
angle between a ray and the longitudinal coordinate
is
\[ \alpha_m=\arctan(p_m/k_m) . \]
 In the limit of small angles and using that  $ k_m\approx\omega/c_s $, it
gives the following SWA condition
\be\label{jan9}
\alpha_m\approx\frac{p_m}{k_m}=\frac{\pi c_s}{\omega y_0}m=\hh m,
\ee
where the inverse number of half-waves in the transverse direction of the
wave-guide
\be\label{ma}
 \hh=\lambda/2y_0
\ee
is the semiclassical parameter.
Transition to the classical or ray dynamics is
therefore the following limits $ \hh\rightarrow 0 $ and
$ m\rightarrow\infty $.

\section{The Ulam map}

In what follows, we present a simple example of a ray dynamics
quantization, where the non-vanishing values of $\hh$ play an important
role. We consider conditions where the limit of small angle propagation
is chaotic.
The ray propagation is characterized by two variables,
namely, the dimensionless longitudinal coordinate $ x $
and the angle (or dimensionless momentum) $ \alpha $. We suppose
that inside the channel a ray undergoes complete refractions from the
boundaries. Therefore the relation between the variables $(\alpha,x)$ for
any two consequent bouncings, for instance, $n$ and $n+1$ (say, from
the bottom boundary) are determined by a geometrical consideration
which is presented in Fig.\ 1a,b.
These relations  form the following map
\bea\label{biz1}
&x_n^*-x_n=(y_0+\vep\phi(x_n^*)/\alpha_n \nonumber \\
&x_{n+1}-x_n^*=(y_0+\vep\phi(x_n^*)/\alpha_{n+1},
\eea
where $\phi(x)$ is the corrugation function introduced for the equation (1)
with an amplitude of modulation $\vep\ll y_0$.
We take into account  the small--angle limit of
(\ref{jan9})
$\tan\alpha\approx\alpha$. The corresponding relation between $\alpha_n$
and $\alpha_{n+1}$ is
\be\label{mb}
\alpha_{n+1}=\alpha_n-2\beta.
\ee
The angle $ \beta $ is defined at the point $ x_n^* $  by a tangent line for
$\phi(x) $ such that $\tan(\beta)=\vep d\phi(x)/dx|_{x=x_n^*}$ (see Fig.\
1b).
Redefining  $z_n\equiv x_n^*$ and neglecting
$\vep\phi(x^*)/y_0=\vep\phi(z)/y_0$ due to the
conditions
$\vep/y_0\ll 1$ and $|\phi(z)|\leq 1$, we obtain from (\ref{biz1}) that
the new relation reads
\be\label{biz2}
z_{n+1}=x_{n+1}+y_0/\alpha_{n+1}=z_n+2y_0/\alpha_{n+1}.
\ee
Eventually, the  following Hamiltonian map describes ray dynamics
\bea\label{biz3}
\alpha_{n+1}&=\alpha_n-2\vep\phi'(z_n) \nonumber \\
z_{n+1}&=z_n+2y_0/\alpha_{n+1},
\eea
where $\phi'(z)\equiv d\phi(z)/dz$. It is simple to see that
$|\prt(\alpha_{n+1},z_{n+1})/\prt(\alpha_n,z_n)|=1$.
In the case when $\phi'(z) $ is periodic, not
necessarily differentiable, the map (\ref{biz3}) is called the Ulam map
\cite{lili1,lili2}.
This map corresponds to a ray propagation when any two nearest bouncings
occur from the opposite boundaries.  The necessary condition  ensures
this effect when the maximal angle determined the corrugation
$ k_0\vep/\pi $ is smaller than the minimum momentum $\alpha_{min}=\hh$.
Hence, the condition of the validity of the map (\ref{biz2}) is
\be\label{reley}
2\vep/\lambda<\pi/k_0y_0.
\ee
In what follows, we consider
\[\phi(z)=\cos(k_0z). \]
For chaotic dynamics the angles are equally distributed after some
time  in some chaotic region of the phase space for any single
initial condition chosen in the chaotic domain (see a phase portrait in
Fig.\ 2). The phase portrait is obtained for a single trajectory
by iteration of the map (\ref{biz3}) when the following
variable change is made: $ u=\alpha/2k_0\vep $ and $ \psi=k_0z $. In this
case, the chaos control parameter is
\be\label{june1}
M=y_0/\vep,
\ee
and it determines the maximal size of the chaotic region
$ |u|<\sqrt{M}$. For the angles, the maximal size of the chaotic region is
$\alpha_{max}\sim 2k_0y_0\sqrt{\vep/y_0} $.

\section{A canonical variable change}

Quantum or wave interference  leads to
localization of chaos and, consequently, to the essential difference
of the initial profile spreading from the ray dynamics \cite{chir}. A
quantum mechanical
counterpart of the map (\ref{biz3}) was the subject of many studies,
related
to the Fermi acceleration mechanism \cite{lili1,sk91,Fermi}.
Some examples of the quantum Fermi acceleration dynamics
have been considered  in the framework of the so--called Generalized
Canonical Transformation \cite{Fermi}.
It should be admitted that the Ulam map is written in the ``energy--time''
canonical variables \cite{lili2}, including the map (\ref{biz3}).
Since the longitudinal coordinates play a role of time, we introduce
the time
parameter in the following dimensionless form $ \nu t=k_0z $, where
$ \nu=k_0y_0 $, while the energy is $ \alpha $, and it scales by 2:
$ \alpha/2\rightarrow\alpha $.
Therefore
the Hamiltonian equations for map
(\ref{biz3}) in the new variables are \bea\label{may1}
&d\alpha/dy=k_0\vep\sin\nu t\sum_{n=-\infty}^{\infty}\delta(y-n)
\nonumber \\
&dt/dy=1/\alpha,
\eea
where the formal time parameter $y$ is the dimensionless transversal
coordinate, which is scaled by $2y_0$. The Hamiltonian is
\be\label{biz5}
H=\ln |\alpha|-\frac{\vep}{y_0}\cos\nu t
\sum_{n=-\infty}^{\infty}\delta(y-n).
\ee
Quantization of the system (\ref{biz5})  is convenient to carry out
in the framework of the transversal momentum and coordinate $(p,y) $
canonical pair \cite{note3,iom,grah}. Moreover, this quantization is
natural.
In the limit $ \alpha\ll 1 $  a simple variable change  could be
suggested. Namely, the Hamiltonian (\ref{biz5}) can be rewritten in a
new form, such that $ t$ will be the real time parameter, while the
dimensionless
transversal coordinate $ y $  and the corresponding transversal momentum
$p$ will be  canonical variables \cite{iom}. Inverting the second equation in
(\ref{may1}) we obtain
\be\label{m2}
\dot{y}=\alpha.
\ee
Now, let us introduce the new unperturbed Hamiltonian $\cH_0(p)$,
such that
\be\label{m3}
\prt\cH_0(p)/\prt p=\alpha,~~\prt\alpha/\prt\cH_0=\alpha.
\ee
It follows from (\ref{m3}) that
\be\label{m4}
\prt\alpha/\prt p=(\prt\alpha/\prt\cH_0)\cdot(\prt\cH_0(p)/\prt p)=\alpha^2.
\ee
Solving (\ref{m3}) and (\ref{m4}) we obtain that
\be\label{m5}
\alpha=-1/p~~\mbox{and}~~\cH_0(p)=\ln|\alpha|=\ln|1/p|.
\ee
This transformation for the unperturbed system corresponds to the mapping
of the interval $\alpha\in(0,\alpha_{max}) $ to the interval
$ p\in(-\infty,-1/\alpha_{max}) $ and the interval
$\alpha\in(-\alpha_{max},0) $ to $ p\in(1/\alpha_{max},\infty) $.
From the first equation in (\ref{may1}) we have
\be\label{m6}
\int_{y_n-0}^{y_{n+1}-0}\frac{d\alpha}{dy}dy=k_0\vep
\int_{y_n-0}^{y_{n+1}-0}\sin\nu t\delta(y-n)dy.
\ee
The integrands from both sides of (\ref{m6}) can be transformed.
Using (\ref{m2}) and (\ref{m4}) we have for the left-side integrand
the following expression
\be\label{m7}
\frac{d\alpha}{dy}=\frac{dt}{dy}\cdot\frac{d\alpha}{dp}\cdot \dot{p}=
\frac{\prt\cH_0}{\prt p}\dot{p}.
\ee
By use (\ref{m2}), the right hand side integral (rhsi) can be rewritten
approximately in the form
\be\label{m8}
\int_{y_n-0}^{y_{n+1}-0}\sin\nu t\delta(y-n)dy=
\int_{t_n-0}^{t_{n+1}-0}\sin\nu t\delta(t-t_n)dt .
\ee
Then taking into account that for the almost equidistant
part of the spectrum, when $p\gg 1$, we have
$ \prt\cH_0(p)/\prt p\approx const $, and  one can write
approximately \cite{note2} that
\be\label{m9}
\frac{d}{dt}\delta(t-t_n)\approx\dot{y}^2
\frac{\prt}{\prt y}\delta(y-n).
\ee
Hence the rhsi in (\ref{m6}) and (\ref{m8}) reads
\bea\label{nov2002}
\int_{y_n-0}^{y_{n+1}-0}\sin\nu t\delta(y-n)dy\approx
-\frac{1}{\nu}\int_{t_n-0}^{t_{n+1}-0}\cos\nu t \frac{d}{d t}
\delta(t-t_n)dt   \nonumber \\
\approx-\frac{1}{\nu}\int_{y_n-0}^{y_{n+1}-0}
\dot{y}\cos\nu t \frac{\prt}{\prt y}\delta(y-n)dy.
\eea
Then we obtain from (\ref{m6})--(\ref{m9}) the following equations
of motion
\bea\label{m10}
\dot{p}=-\frac{k_0\vep}{\nu}\cos\nu t
\sum_{n=-\infty}^{\infty}\frac{\prt}{\prt y}\delta(y-n)
~~\mbox{and}~~\dot{y}=-1/p
\eea
with the Hamiltonian
\be\label{m11}
\cH=-\ln|p|+\frac{k_0\vep}{\nu}\cos\nu t
\sum_{n=-\infty}^{\infty}\delta(y-n)=\cH_0(p)+V(y,t).
\ee

\section{Quantization of the Ulam map}

A quantization procedure now corresponds to the consideration of
the Shr\"o\-di\-nger equation for the wave function. The
semiclassical consideration requires
that the width of the perturbative potential $V$ in (\ref{m11}) is larger
than the wavelength. Therefore, it is necessary to restrict the summation
in the Fourier expansion of the $ \delta_{2\pi}(\theta) $ potential. Thus
we
obtain a new potential with the width of a spike equaled to $ 2\pi/N $:
\[
\delta_N(\theta)=2\sum_{k=0}^N\cos k\theta -1\equiv\sum_{k=-N}^N
\exp{ik\theta},
\]
where $ \delta_N(\theta) $ tends to $ \delta_{2\pi}(\theta) $ at $ N $
tends to infinity.

\subsection{Floquet theory}

Since the Hamiltonian (\ref{m11}) is periodic in time, the Floquet theory
is used in what follows. In this case the wave function due to the Floquet
theorem is
\[ \psi(y,t)=e^{-i\Lambda t}\psi_{\Lambda}(y,t) , \]
where $ \psi_{\Lambda}(y,t+2\pi/\nu)=\psi_{\Lambda}(y,t) $ is the periodic
eigenfunction with the period of the perturbation.
The Schr\"odinger equation
\be\label{m12}
i\hh\frac{\prt}{\prt t}\psi(y,t)=\hat{\cH}(p,y,t)\psi(y,t)
\ee
corresponds to the eigenvalue problem for the Floquet
operator $ \hat{F} $
\be\label{m13}
\hat{F}\psi_{\Lambda}=\hh\Lambda\psi_{\Lambda}.
\ee
The Floquet operator is
\be\label{m14}
\hat{F}=-i\hh\frac{\prt}{\prt t}+\hat{\cH}(p,y,t)=
\hat{F}_0+\hat{V}(y,t).
\ee
The solution of (\ref{m13}) is considered as a superposition of the
unperturbed basis
\be\label{m15}
 |n,j\rgl\equiv|n\rgl|j\rgl =\sqrt{\frac{\nu}{2\pi}}
e^{i2\pi ny}e^{-ij\nu t},
\ee
which is the eigenfunction of the
unperturbed system $ \hat{F}_0=-i\hh\prt/\prt t+\hat{\cH}_0 $ and
$ p|n\rgl=\hh n|n\rgl $.
Taking into account that the perturbation is periodic in $y$,
we obtain for the eigenfunction
\be\label{m16}
\psi_{\Lambda}(y,t)\equiv|\psi_{\Lambda}(y,t)\rgl=e^{-i\theta y}
\sum_{n,j}\phi_{\Lambda,\theta}(n,j)|n,j\rgl,
\ee
where $\theta$ is a quasi-momentum.
The coefficients of the expansion
$\phi_{n,j}\equiv\phi_{\Lambda,\theta}(n,j)=
\lgl j,n|\psi_{\Lambda}(y,t)\rgl $ can be found from the following
equation
\be\label{m17}
\left[\clE(n+\theta)-\hh\nu j\right]\phi_{n,j}-
\frac{k_0\vep}{2\nu}\sum_{n'}\left[\phi_{n',j+1}+
\phi_{n',j-1}\right]=\hh\Lambda\phi_{n,j},
\ee
where $ \clE(n+\theta)=\lgl n|\hat{\cH_0}(p+\theta)|n\rgl $ and
$ \sum_{n'}\phi_{n',j}<\infty $.

\subsection{An exact solution in the linear approximation}

Some simplification of the equation can be made. Taking into account that
in the range of the almost equidistant spectrum for $p_0\gg 1$, one can
consider
approximately that $ \clE(p_0+\theta+\hh n)\approx\clE_0+\hh v(p_0)n $,
where $ \clE_0\equiv \clE(p_0+\theta) $ and
$ v(p_0)\equiv\frac{d}{dp_0}\clE(p_0+\theta) $.

To lift the summation over $n'$ in (\ref{m17}), a $\sinc$-function
$ \sinc x=\frac{\sin x}{x} $ is introduced. It possesses $\delta$-like
properties
\be\label{m18}
\sum_m\sinc\pi(m-a) =1,~~\mbox{and}~~ \sinc 0=1.
\ee
Then the solution of (\ref{m17}) is cast in the form of the
sinc-function
\be\label{m19}
\phi_{n,j}=Z_j\sinc\pi[n-(\Lambda-\clE_0)/v],
\ee
where coefficients $ Z_j $ correspond to the following equation
\bea\label{m20}
\left[\hh vn-\hh\nu j-\hh(\Lambda-\clE_0/\hh)\right]
Z_j\sinc\pi[n-(\Lambda-\clE_0)/v] \nonumber \\
+ \frac{k_0\vep}{2\nu}\left[Z_{j+1}+Z_{j-1}\right]=0.
\eea
Counting for the fixed $n$ and $j$ that the quasienergy spectrum is
\be\label{m21}
\hh\Lambda=\clE_0+\hh nv,
\ee
we obtain the following relation for the Bessel functions \cite{bes}
\be\label{m22}
2jZ_j=\frac{k_0\vep}{\hh\nu}[Z_{j+1}+Z_{j-1}]
\ee
with the solution $ Z_j=J_j(\frac{k_0\vep}{\hh\nu})=
J_j(\frac{2\vep}{\lambda}) $. In the case when $k_0y_0\gg 1$,
the effective number of transitions due to the perturbation is restricted
by the relation for the Bessel function of a small argument
$ 2\vep/\lambda \ll 1$ \cite{bes}, such that
\be\label{july1}
J_j(\frac{2\vep}{\lambda})\sim (\frac{2\vep}{\lambda})^j=
\exp\{-j\ln(\lambda/2\vep)\}.
\ee
Therefore, the effective number of interacting eigenmodes is restricted by
the relation
\be\label{m23a}
 |j|<j_0\sim 1/\ln(\lambda/2\vep).
\ee
In the opposite case when the period of the corrugation is larger than
the transversal size of the channel, one obtains from (\ref{reley}) that
$ 2\vep/\lambda \gg 1 $. In this case the effective number of
transitions due to the perturbation is restricted by the relation
\be\label{m23}
 |j|<j_0\sim 2\vep/\lambda,
\ee
since the Bessel functions in this case decay faster than the exponential
when $ j>j_0 $ \cite{bes}. One should also recognize that in a general case
the approximation for the quasi--equidistant spectrum could be not valid.
Then the approximate analysis should be carried out beyond the
linear approximation used here for the expression
(\ref{m23}) \cite{iom}.

\section{Conclusion}

The ratio between the wavelength and the amplitude of the corrugation
is important not only for the validity of the Ulam map (\ref{biz3}), but
because it also determines the relation on the ray chaos for the quantum
or wave process. Since chaotic ray dynamics is strongly damped by
quantum effects,  the quantum system behaves like its classical
counterpart of eq. (\ref{biz3}) on a finite time scale only. Therefore,
one can discuss the chaos localization phenomenon by quantum effects.
It means that quantum chaotic dynamics is localized in some energy scales
determined by (\ref{m23a}) and (\ref{m23}). Borrowing some familiar
terminology, one defines this range as a localization length.
In the units of the unperturbed spectrum, the localization length is
$\Delta n<j_0\nu/v $. For the angles it reads
\be\label{m24}
\Delta\alpha_q=\frac{1}{p_0}-\frac{1}{(p_0+\hh\Delta n)}\approx
\left\{ \begin{array}{r@{\quad:\quad}l}
(k_0\vep)\cdot\alpha_0 & \lambda\ll\vep  \\
(k_0\vep)\cdot\frac{\lambda}{2\vep}\cdot
\left[\ln(\frac{\lambda}{2\vep})\right]^{-1}\cdot\alpha_0
&
\lambda\gg\vep \, ,\end{array}\right.
\ee
and the angles' width $ \Delta\alpha_q $ also defines the width of the
spreading wave packet.
We cannot but to admit two important characteristics (or
parameters)  which determine the localization lengths in (\ref{m24}).
The first one
follows from (\ref{ma}),(\ref{june1}) and (\ref{m23}). The semiclassical
parameter $\hh$, the localization length $ 2\vep/\lambda $ and the chaos
control parameter $M$ form the following relation, familiar in quantum
chaos \cite{bz}
\be\label{m25}
1/M=(\lambda/2y_0)\cdot(2\vep/\lambda) .
\ee
The conditions of quantum chaos are $ \lambda/y_0\ll 1$ and $ M\gg 1 $.
This interplay of classical or ray chaos
with the quantum interference is an important mechanism of quantum chaos
or quantum localization of classical chaos.
The second one is concerned with the so--called Rayleigh condition
$k_0\vep<0.38$ \cite{rayleigh} imposed on the system for a spreading wave.
In our case $ k_0\vep <\hh\ll 1 $, which follows from (\ref{reley}).
One may also understand this result from a point of view
of wave dynamics. The excitation of an initial wave with an arbitrary
angle $\alpha_0$ means that an initial wave packet is a superposition of
eigenmodes of $\alpha_m$ (see (\ref{jan9})). If the corrugation is weak
enough, the wave packet is a narrow superposition of the Floquet
eigenfunctions of (\ref{jan4}) with $ \alpha_m $ being close to the
initial angle $\alpha_0$, and its width $\Delta\alpha_q$ remaining a
constant value during the evolution. An estimation of this width is
determined by Eq. (\ref{m24}):  $|\alpha_m-\alpha_0|\leq\Delta\alpha_q$.
We should stress that the present analysis can be considered as a
qualitative estimation of this width as well.

We thank Prof. George Zaslavsky for numerous stimulating discussions.
We thank Ann Pitt for her help in preparing this paper.
A.I. also uses this opportunity to thank Prof. George Zaslavsky for his
hospitality at the Courant Institute where part of the work was done.
This research was supported  by the Minerva Center of Nonlinear Physics
of Complex Systems.

\newpage
\section*{Figure captions}

\noindent  Fig. 1. A ray dynamics sketch for constructing map (\ref{mb}):
$\alpha^{\prime}=\alpha-2\beta$.

\noindent  Fig. 2. Phase portrait of the map (\ref{biz3}) for a single
trajectory with an initial condition $(\psi,u)=(0,0.1) $. The following
variable change is made $ u=\alpha/2k_0\vep $ and  $ \psi=k_0z $, while
the chaos control parameter  is $M=y_0/\vep$. The trajectory is shown after
125000 iterations for $ M=2\pi\cdot 125 $.

\newpage
\begin{figure}
\begin{center}
\epsfxsize=7.6cm
\leavevmode
    \epsffile{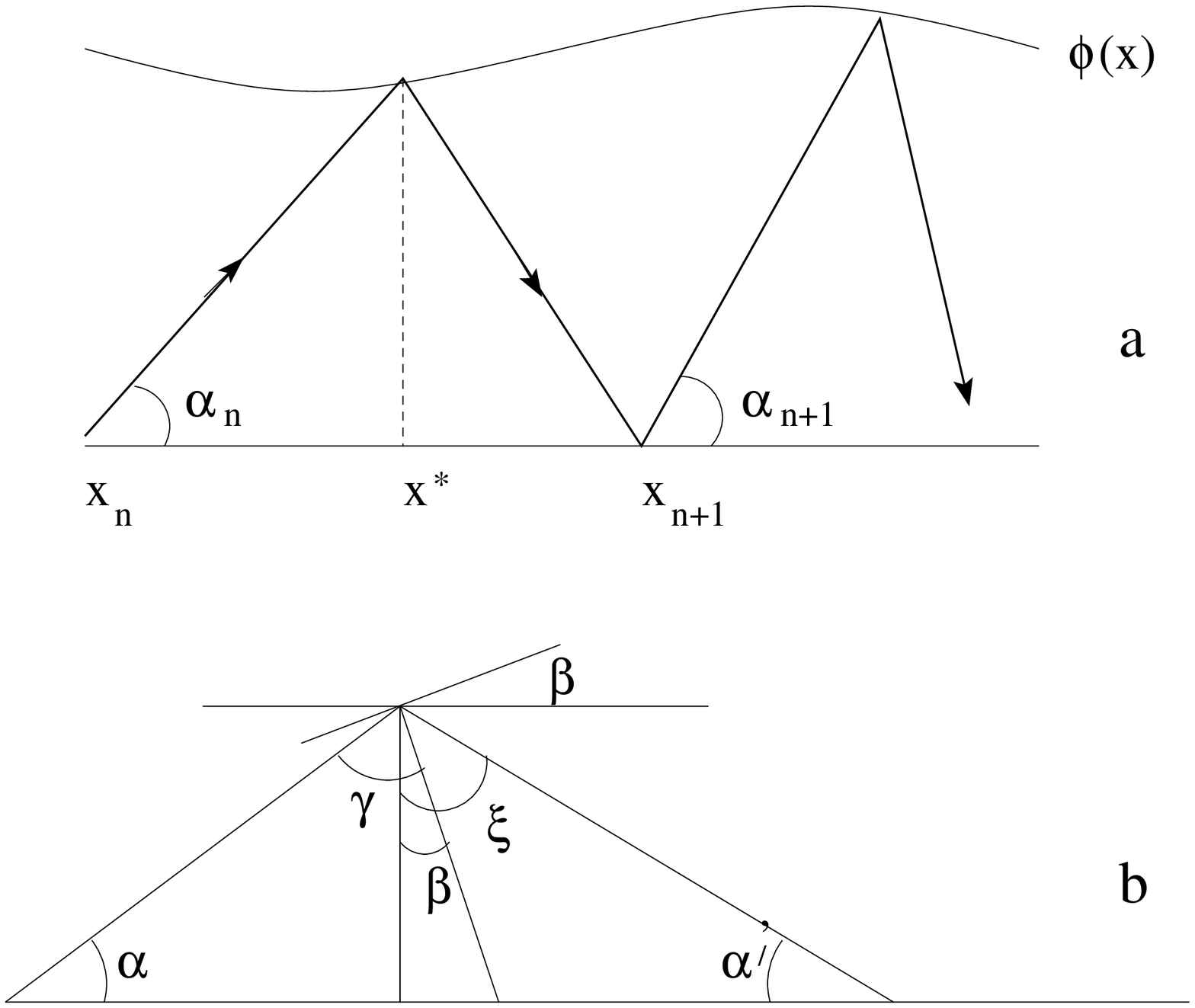}

\caption{}
\label{fig:????}
\end{center}
\end{figure}
\newpage
\begin{figure}
\begin{center}
\epsfxsize=7.6cm
\leavevmode
    \epsffile{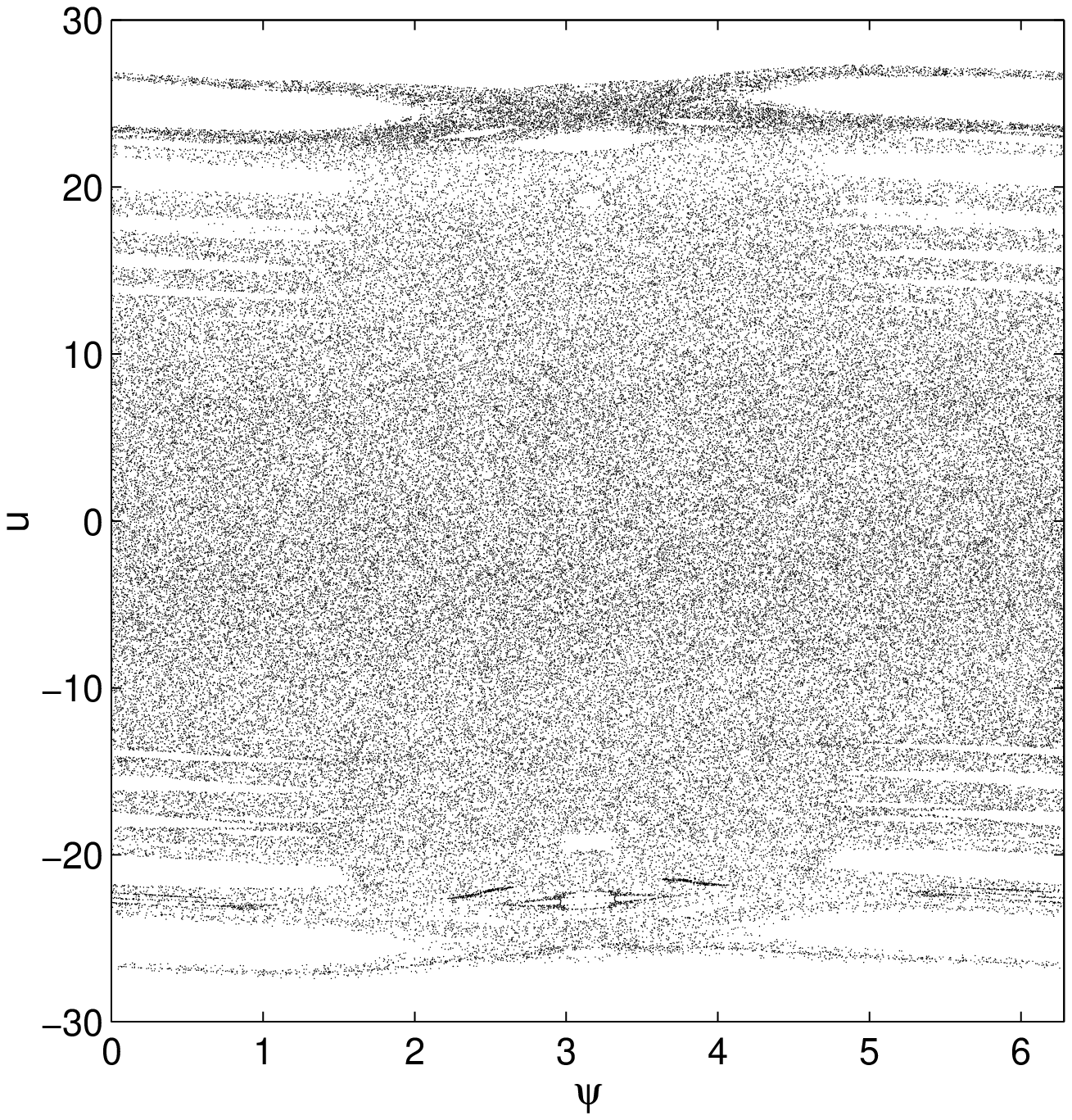}
\caption{}
\label{fig:??????}
\end{center}
\end{figure}


\begin{thebibliography}{99}

\bibitem{tappert} F.D. Tappert, in Wave Propagation and Underwater
Acoustics , edited by J.B. Keller and J.S. Papadakis, Springer-Verlag,
Berlin 1977, p. 224.

\bibitem{jkps94} F.B. Jensen, W.A. Kuperman, M.B. Porter, and H. Schmidt,
Computational Ocean Acoustics, AIP, Woodbury, NY 1994.

\bibitem{vz00} A.L. Virovlyansky and G.M. Zaslavsky, Chaos 10 (2000) 211.

\bibitem{mw79} W. Munk and C. Wunsch, Deep--Sea Res. 26 (1979) 123.

\bibitem{atoc}  The ATOC Consortium, Science 1326 (1988).

\bibitem{8a} M.G. Brown , F.D. Tappert, and G. Goni, Wave Motion 14
(1991) 93.

\bibitem{pbtb88} D.R. Palmer, M.G.Brown, F.D. Tappert, and M.F. Bezdek,
Geophys. Res. Lett. 15 (1988) 569.

\bibitem{sbt92} K.B. Smith, M.G. Brown, and F.D. Tappert, J. Acoust.
 Soc. Am.  91 (1992) 1939, 1950.

\bibitem{sfw97} J. Simmen, S.M. Flatte, and G.-Y. Wang, ibid 102 (1997)
239.

\bibitem{az91} S.S. Abdullaev and G.M. Zaslavsky, Sov. Phys. Usp. 34
(1991) 645.

\bibitem{sz99} B. Sundaram and G.M. Zaslavsky, Chaos 9 (1999) 483.

\bibitem{svz01} I.P. Smirnov, A.L. Virovlyansky, and G.M. Zaslavsky,
Phys. Rev. E 64(2001) 036221.

\bibitem{gutzwiller} M.C. Gutzwiller, Chaos in Classical and Quantum
Mechanics, Springer-Verlag, New York, 1990.

\bibitem{12a} G.M. Zaslavsky, Phys. Rep. 80 (1981) 157.

\bibitem{12b} G.M. Zaslavsky and S.S. Abdullaev, Chaos 7 (1997) 182.

\bibitem{bz1978} G.P. Berman and G.M. Zaslavsky, Physica A 91 (1978) 450.

\bibitem{bz} G.P. Berman and G.M. Zaslavsky, Physica A 111 (1982) 17.

\bibitem{note1} It should be noted that considerating the higher orders of
$\vep$ in the perturbation approach does not change the form of the
dispersion relation (\ref{jan6}).
It leads to the correction of the order of $\vep^2$,
except for those regions of $k$, where an interaction between
different modes (say $ \omega_m(k)$ and $ \omega_{m'}(k) $)
leads to the avoiding modes crossing, and the corrections to the
dispersion law in this case is of the order of $\vep$.

\bibitem{lili1} A.J. Lichtenberg and M.A. Liberman, Physica D 1 (1980)
291.

\bibitem{lili2} A.J. Lichtenberg and M.A. Liberman, Regular and
Stochastic Motion, Springer-Verlag, New York, 1983.

\bibitem{chir} G. Casati and B.V. Chirikov in Quantum Chaos:
Between Order and Disorder, edited by G. Casati and B.V. Chirikov,
Cambridge, 1995.

\bibitem{Fermi} A. Munier, J.R. Burgan, M. Feix, and E. Fijalkov,
J. Math. Phys.  22 (1981) 1219; J.V. Jose and R. Gordery,
Phys. Rev. Lett. 56 (1986) 290; C. Scheininger and M. Kleber,
Physica D 50 (1990) 391.

\bibitem{sk91} C. Scheininger and M. Kleber, Physica D  50, (1990) 391.

\bibitem{note3} Semiclassical quantization of the system in the framework
of the energy--time canonical variables is not a properly defined  task,
because the operator $ \hat{\alpha}=-i\hh\prt/\prt t $ is not defined
semiclassically \cite{iom}.
Another deficiency is an appearance of the non-physical time
($y$-coordinate in our case) for a wave function. It has been already
pointed out \cite{grah,iom} that in this case it is impossible to study
quantum dynamics.
Another question relates to the quasienergy spectrum which is false,
as can seen here.

\bibitem{iom} A. Iomin, S. Fishman, and G.M. Zaslavsky, Semiclassical
quantization of maps with a variable time scale, unpublished.

\bibitem{grah} R. Graham, Europhys. Lett. 7 (1988) 671.

\bibitem{note2}
We restrict ourselves by the expression (\ref{m9}), since
an exact analysis is quite cumbersome (see \cite{iom}) and is not within
the scope of the present analysis.

\bibitem{bes} E. Jahnke, F. Emde, F. L\"osh, Tables of a Higher
Functions, New York, McGrow-Hill, 1960.

\bibitem{rayleigh} L. Rayleigh, Proc. Ray. Soc. A  79 (1907) 399.

\end{thebibliography}
\end{document}